\documentstyle[11pt,newpasp,twoside]{article}
\index{pulsars!ages} \index{pulsars!associations with supernova
remnants} \index{pulsars!B1757--24} \index{pulsars!J1747--2958}
\index{pulsars!displacement from center of supernova remnants}
\index{pulsars!proper motion} \index{pulsars!velocities}
\index{pulsar wind nebulae} \index{pulsar wind
nebulae!G359.23--0.82} \index{radio source!G5.27--0.90}
\index{supernova remnants} \index{supernova remnants!associations
with pulsars} \index{supernova remnants!G5.4--1.2}
\index{supernova remnants!shell}

\def\edcomment#1{\iffalse\marginpar{\raggedright\sl#1\/}\else\relax\fi}
\reversemarginpar

\begin{document}
\title{The Association of PSR B\,1757$-$24 and the SNR G\,5.4$-$1.2}
 \author{V.V.Gvaramadze}
\affil{Sternberg Astronomical Institute, Moscow State University,
Universitetskij Pr.~13, Moscow, 119992, Russia;}
\affil{E.K.Kharadze Abastumani Astrophysical Observatory, Georgian
Academy of Sciences, A.Kazbegi ave.~2A, Tbilisi, 380060, Georgia}

\begin{abstract}

The association of PSR B\,1757$-$24 and the supernova remnant
(SNR) G\,5.4$-$1.2 was recently questioned by Thorsett et al.
(2002) on the basis of proper motion measurements of the pulsar
and the ``incorrect" orientation of the vector of pulsar
transverse velocity [inferred from the orientation of the
cometary-shaped pulsar wind nebula (PWN)]. We showed, however,
that the association could be real if both objects are the
remnants of an off-centred cavity supernova (SN) explosion.
\end{abstract}

\section{Introduction}

Recent proper motion measurements of PSR B\,1757$-$24 by Thorsett,
Brisken, \& Goss (2002) put a $2\sigma$ upper limit on the pulsar
transverse velocity,  $v_{\rm p} \leq 160 \, d_{5} \,{\rm
km}\,{\rm s}^{-1}$, where $d_{5}$ is the distance to the pulsar in
units of 5 kpc. This upper limit is at least an order of magnitude
less than the velocity estimate inferred from the angular
displacement of PSR B\,1757$-$24 from the geometric centre of
G\,5.4$-$1.2 (Frail \& Kulkarni 1991; Manchester et al. 1991).
Thorsett et al. interpreted the discrepancy between the
``measured" and inferred velocities as an indication of equally
large discrepancy between the kinematic age of the system, $t_{\rm
kin} =l/v_{\rm p}$, where $l$ is the distance traveled by the
pulsar from its birthplace, and the characteristic age of the
pulsar, $\tau = P/(n-1)\dot{P}$. The latter discrepancy and the
``incorrect" orientation of the inferred line of pulsar proper
motion (the cometary-shaped PWN does not point to the geometric
centre of G\,5.4$-$1.2; Frail, Kassim \& Weiler 1994) constitutes
two arguments against the physical association of PSR B\,1757$-$24
and G\,5.4$-$1.2 proposed by Thorsett et al. (2002). In this paper
we show, however, that the association could be real if both
objects are the remnants of a SN explosion within a bubble
blown-up by the moving SN progenitor star during the Wolf-Rayet
(WR) phase of its evolution.

\section{SNR G\,5.4$-$1.2 and its  progenitor star}

Let us explain why we believe that the pre-SN was a WR star and
that the SN exploded within the WR bubble, but not in the bubble
created during the preceding main-sequence (MS) phase. In our
reasoning we proceed from the fact that a young neutron star (born
with a moderate kick velocity of appropriate orientation) can
overrun the shell of the associated SNR only on conditions that:
a) the SN exploded within a pre-existing bubble surrounded by a
massive shell, b) the SN explosion site was significantly offset
from the centre of the bubble (e.g. Gvaramadze 2002a,b). It is
unlikely, however, that these conditions can be fulfilled for the
MS bubbles. Indeed, simple estimates show that most of massive
stars explode outside their MS bubbles, while the bubbles stall
and loss their shells well before the end of the MS phase
(Brighenti \& D'Ercole 1994). On the other hand, if a massive star
ended its evolution as a WR star, the energetic WR wind could
create a new large-scale bubble, whose supersonic expansion drives
a shell of swept-up interstellar matter (ISM) during the whole
relatively short WR phase. Besides, it is the short duration of
the WR phase that results in that even a runaway massive star
could explode within its WR bubble.

We assume that the SN exploded near the west edge of the WR bubble
(cf. Gvaramadze \& Vikhlinin 2003) on the line defined by the
cometary-shaped PWN, i.e. the SN exploded $\simeq 9 \, d_5$ pc
east of the current position of the pulsar (or about $3.5 \, d_5$
pc behind the west edge of G\,5.4$-$1.2). In this case, $t_{\rm
kin} (\simeq 5.4\times 10^4$ yr) could be reconciled with $\tau$
if $n\leq 1.6$, i.e. for braking indices comparable with $n$
measured for the Vela pulsar (cf. Thorsett et al. 2002).

The further evolution of the blast wave depends on the mass of the
pre-existing shell, $M_{\rm sh}  = (4\pi / 3) R_{\rm sh} ^3  \rho
_{\rm ISM}$, where $R_{\rm sh}$ is the radius of the shell, $\rho
_{\rm ISM} =1.3 m_{\rm H} n_{\rm ISM}$, $m_{\rm H}$ is the mass of
a hydrogen atom and $n_{\rm ISM}$ is the number density of the
ambient ISM. The number density could be evaluated by comparing
the observed minimum size of the PWN ahead of the moving pulsar,
$R_{\rm n}$, with the theoretically predicted one, $\kappa R_{0} =
\kappa (|\dot{E}| /4\pi c \rho _{\rm ISM} v_{\rm p} ^2 )^{1/2}$,
where $\kappa \simeq 1.26$ (Bucciantini 2002), $R_{0}$ is the
stand-off distance, $|\dot{E}| \simeq 2.6 \times 10^{36} \, {\rm
erg} \, {\rm s}^{-1}$ is the spin-down luminosity of the pulsar
and $c$ is the speed of light. For $R_{\rm n} = 3.6\times 10^{-2}
d_{5}$ pc (Gaensler \& Frail 2000) and $v_{\rm p} =160 \, {\rm
km}\,{\rm s}^{-1}$, one has $n_{\rm ISM} \simeq 1.0 \, {\rm
cm}^{-3}$. Then assuming that $R_{\rm sh} =20$ pc, one has $M_{\rm
sh} \simeq 10^3 \, M_{\odot}$.

The numerical simulation of cavity SN explosions by Tenorio-Tagle
et al. (1991) showed that the SN blast wave evolves into a
momentum-conserving stage if the mass of the pre-existing shell
was larger than $\simeq 50 M_{\rm ej}$, where $M_{\rm ej}$ is the
mass of the SN ejecta. For any reasonable initial mass of the SN
progenitor, one has that $M_{\rm sh} >> 50\, M_{\rm ej}$. Thus the
SNR G\,5.4$-$1.2 is in the radiative stage (with the initial
expansion velocity of $\simeq 100 \, {\rm km} \, {\rm s}^{-1}$),
so that the pulsar can easily overrun the SNR.

\section{G\,5.27$-$0.9}

We now discuss the origin of a compact
source G\,5.27$-$0.9 located between PSR B\,1757$-$24 and
G\,5.4$-$1.2 (e.g. Frail \& Kulkarni 1991).
We suggest that G\,5.27$-$0.9 is a lobe of a low
Mach number jet of
gas outflowing from the interior of G\,5.4$-$1.2 through the hole
bored in the SNR's shell by the escaping pulsar.

The gas velocity at the origin of the jet is $v_{\rm j} \simeq \sqrt
{3} \, c_{\rm j}$, where $c_{\rm j}$ is the sound speed of the
outflowing gas. The structure and the dynamics of
supersonic jets propagating through the ambient medium are mainly
determined by two parameters:
the jet Mach number, ${\cal{M}}_{\rm j} =v_{\rm j} /c_{\rm j}$,
and the jet to ambient medium density ratio, ${\rho}_{\rm j} /\rho
_{\rm ISM}$ (see Norman et al. 1982). In our case
${\cal{M}}_{\rm j} \simeq 1.7$ and ${\rho}_{\rm j} /\rho _{\rm
ISM} << 1$.
Numerical simulations conducted by Norman et al. (1982) showed
that a low-density, Mach 1.5 jet ends itself in a gradually inflating and
slowly-moving lobe. The morphological similarity of this lobe (see
Fig.\,10a of Norman et al. 1982) and G\,5.27$-$0.9
(see Fig.\,1b of Frail \& Kulkarni 1991) allows us
to consider the existence of inner bright spots in G\,5.27$-$0.9
and the edge-darkened appearance of this source
as indications that the jet is already
reached its maximum spatial extent (see Norman et al. 1982).
Therefore the pulsar moving along the jet axis was able to overrun
the lobe and now it travels through the ISM.

\section{PSR B\,1757$-$24 and its PWN}

Is is clear that the proper motion of a neutron star born in an
off-centred cavity SN explosion could be oriented arbitrarily with
respect to the geometric centre of the associated SNR (Gvaramadze
2002a,b). Therefore one should not comment why the cometary-shaped
PWN does not point to the geometric centre of G\,5.4$-$1.2. Let us
briefly discuss some points related to the origin of this nebula.

The supersonic motion of PSR B\,1757$-$24 through the ISM results
in the origin of an elongated structure, where the pulsar wind is
swept back by the ram pressure. The region occupied by the wind is
bounded by a contact discontinuity, which asymptotically becomes
cylindrical with a characteristic radius $R\simeq 0.85
{\cal{M}}_{\rm p} ^{3/4} (1-0.85 {\cal{M}}_{\rm p} ^{-1/2}
)^{-1/4}  R_0$, where ${\cal{M}}_{\rm p} = v_{\rm p} /c_{\rm ISM}$
and $c_{\rm ISM}$ is the sound speed of the ambient ISM
(Bucciantini 2002). For the temperature of the ambient ISM of
$\simeq 8\,000 \, {\rm K}$, one has that $R (\simeq 7^{\arcsec} \,
d_5 ^{-1})$ is few times larger than the half-width of the PWN,
i.e.  most of the pulsar wind is unobservable.

We suggest that the non-thermal X-ray emission of the cometary
tail behind the pulsar (Kaspi et al. 2001) is due to the
synchrotron losses of the relativistic pulsar wind shocked at the
termination shock, which extends in the tail up to a distance of
$L \simeq 1.29 \, {\cal{M}}_{\rm p} R_0$ (see Bucciantini 2002 and
Fig.\,1 therein) and where the wind particles acquire non-zero
pitch angles. An indirect support to this suggestion comes from
the comparison of $L \simeq 19^{\arcsec} \, d_5 ^{-1}$ with the
observed length of the X-ray tail of $\simeq 20^{\arcsec}$.

We also suggest that the (non-thermal) radio emission of the PWN
originates in the vicinity of the termination shock and in a much
more extended narrow cylindrical region of subsonically moving
shocked pulsar wind (cf. Bucciantini 2002). This suggetion implies
that in the absence of the radio source G\,5.27$-$0.9 the radio
tail would be much longer than its X-ray counterpart [perhaps as
long as the tail of the radio nebula ``Mouse" (G\,359.23$-$0.82;
Yusef-Zadeh \& Bally 1987) powered by a young pulsar PSR
J\,1747$-$2958 (whose spin characteristics are almost the same as
those of PSR B\,1757$-$24; Camilo et al. 2002)].

\section{Concluding remark}

To conclude,  we note that the idea  of off-centred cavity SN
explosion could be used not only to assess the reliability of
proposed neutron star/SNR associations (Gvaramadze 2002a; Bock \&
Gvaramadze 2002),  but also to explain the diverse morphologies of
the known SNRs (Gvaramadze 2002b, 2003; Gvaramadze \& Vikhlinin
2003) and to search for new stellar remnants associated with SNRs
(Gvaramadze \& Vikhlinin 2003).

\acknowledgments
I am grateful to R.N.Manchester, M.Orine and H. Rickman, whose
support allows me to attend the symposium.

\end{document}